\documentclass[12pt,a4paper,notoc]{JHEP}

\usepackage{amsmath,amsfonts}
\usepackage{epic,eepic}

\newcommand{\al}{\alpha}
\newcommand{\lam}{\lambda}
\newcommand{\half}{\frac{1}{2}}
\newcommand{\eq}[1]{eq.(\ref{#1})}

\newcommand{\bphi}{\boldsymbol{\phi}}
\newcommand{\weight}{\boldsymbol{\lambda}}
\newcommand{\aroot}{\boldsymbol{\alpha}}
\newcommand{\real}[1]{\text{Re}(#1)}
\newcommand{\imaginary}[1]{\text{Im}(#1)}

\newcommand{\reals}{\mathbb{R}}
\newcommand{\lie}{g}
\newcommand{\kac}{\eta}
\newcommand{\one}[1]{\stackrel{1}{#1}}
\newcommand{\two}[1]{\stackrel{2}{#1}}
\newcommand{\ot}[1]{{#1}}
\newcommand{\ct}{{\cal{T}}}
\newcommand{\cct}{\tau}
\newcommand{\tr}{\text{tr}}

\newcommand{\invo}{\sigma}
\newcommand{\bH}{{\bf{H}}}

\newtheorem{proposition}{Proposition}

\newcommand{\drawcenteredtext}[3]{\put(#1,#2){\makebox(0,0){#3}}}%

\newcommand{\drawpath}[4]{\path(#1,#2)(#3,#4)}%
\newcommand{\drawvector}[5]{\put(#1,#2){\vector(#4,#5){#3}}}%

\newcommand{\drawleftbrace}[3]%
{\drawcenteredtext{#1}{#2}{$\left\{ \rule[0mm]{0mm}{#3mm} \right.$}}%

\newcommand{\drawrightbrace}[3]%
{\drawcenteredtext{#1}{#2}{$\left\} \rule[0mm]{0mm}{#3mm} \right.$}}%

\newcommand{\drawoverbrace}[3]%
{\drawcenteredtext{#1}{#2}{$\overbrace{\rule[0mm]{#3mm}{0mm}}$}}%

\newcommand{\drawunderbrace}[3]%
{\drawcenteredtext{#1}{#2}{$\underbrace{\rule[0mm]{#3mm}{0mm}}$}}%

\newcommand{\drawarc}[5]%
{\put(#1,#2){\arc{#3}{#4}{#5}}}%

\title{Soliton-preserving boundary condition in affine Toda field theories}

\author{Gustav W Delius \\
Department of Mathematics, King's College London,\\
Strand, London WC2R 2LS, UK.\\
E-mail: \email{delius@mth.kcl.ac.uk}\\
\href{http://www.mth.kcl.ac.uk/~delius/}
{WWW: \tt{http://www.mth.kcl.ac.uk/$\sim$delius/}}
}

\abstract{We give a new integrable boundary condition in
affine Toda theory which is soliton-preserving in the
sense that a soliton hitting the boundary is reflected
as a soliton. All previously known integrable boundary 
conditions forced a soliton to be converted into an 
antisoliton upon reflection. We prove integrability
of our boundary condition using a generalization of
Sklyanin's formalism.
}

\begin{document}

\section{Introduction}

Associated to every complex affine Lie algebra $\hat{\lie}$ there is an
integrable relativistic field theory in 1+1 dimensions, 
called affine Toda
theory, described by the equations of motion
\begin{equation}
\ddot{\bphi}-\bphi^{\prime \prime }+m^2\sum_{i=0}^n\kac _i%
\aroot_ie^{\aroot_i\cdot \bphi}=0.
\label{phieom}
\end{equation}
Here $\bphi$ is an n-component bosonic field, where $n$ is the rank
of $\hat{g}$. $\aroot_i$, $i=1,\dots,n$ are the simple roots of
the finite dimensional Lie algebra $\lie$ underlying $\hat{\lie}$ and
$\aroot_0=-\sum_{i=1}^n\kac_i\aroot_i$ is the extra simple root that
needs to be added to obtain the extended Dynkin diagram of $\hat{g}$.
We have defined $\kac_0=1$. $m$ is a mass parameter and we will choose
units so that $c=m=1$ throughout this letter. We have rescaled the
fields so that the coupling constant $\beta$ of affine Toda theory 
does not enter the equations of motion. The simplest affine Toda theory
is that for $\lie=sl(2)$ which is the sinh-Gordon theory (or the sine-Gordon 
theory after rescaling the field by a factor of $i$). For a review
of affine Toda theory see for example \cite{Cor94}.

The integrability of affine Toda theories manifests itself in the
existence of soliton solutions \cite{Hol92,Oli93b}. These are kink configurations
interpolating  between different minima of the potential.
The potential corresponding to the equations of motion \eqref{phieom}
is
$
V[\bphi]=\sum_{i=0}^n\eta_i\left(e^{\aroot_i\cdot\bphi}-1\right)
$
and its degenerate minima are at $\bphi=2\pi i\weight$ where 
$\weight$ are the coweights of $\lie$.
The special property of solitons is that when two of them meet they
pass through each other and reemerge with their original shape.
In the quantum theory the solitons give 
rise to particle
states and their S-matrices have been determined \cite{Hol93,Gan96}.

\FIGURE{
\setlength{\unitlength}{0.6mm}
\begin{picture}(112,40)
\thicklines
\path(8.0,24.0)(8.0,24.0)(8.29,24.0)(8.57,24.0)(8.86,24.02)(9.12,24.02)(9.38,24.04)(9.64,24.08)(9.89,24.11)(10.13,24.13)
\path(10.13,24.13)(10.37,24.18)(10.6,24.22)(10.81,24.25)(11.04,24.31)(11.25,24.36)(11.45,24.41)(11.63,24.47)(11.82,24.54)(12.02,24.61)
\path(12.02,24.61)(12.2,24.68)(12.37,24.75)(12.54,24.83)(12.7,24.9)(12.86,24.99)(13.01,25.06)(13.15,25.15)(13.3,25.25)(13.45,25.34)
\path(13.45,25.34)(13.57,25.43)(13.71,25.52)(13.85,25.61)(13.96,25.72)(14.09,25.81)(14.2,25.93)(14.32,26.02)(14.44,26.13)(14.54,26.25)
\path(14.54,26.25)(14.65,26.36)(14.76,26.47)(14.87,26.58)(14.96,26.7)(15.06,26.81)(15.15,26.93)(15.26,27.04)(15.36,27.15)(15.45,27.27)
\path(15.45,27.27)(15.54,27.4)(15.62,27.52)(15.71,27.63)(15.8,27.75)(15.89,27.88)(15.98,28.0)(16.09,28.11)(16.18,28.22)(16.27,28.34)
\path(16.27,28.34)(16.36,28.47)(16.45,28.59)(16.54,28.7)(16.62,28.83)(16.72,28.95)(16.82,29.06)(16.92,29.18)(17.02,29.29)(17.12,29.4)
\path(17.12,29.4)(17.22,29.52)(17.32,29.63)(17.44,29.74)(17.54,29.84)(17.65,29.95)(17.78,30.06)(17.89,30.16)(18.02,30.27)(18.13,30.36)
\path(18.13,30.36)(18.27,30.45)(18.4,30.56)(18.54,30.65)(18.68,30.74)(18.82,30.83)(18.97,30.91)(19.12,31.0)(19.29,31.08)(19.45,31.15)
\path(19.45,31.15)(19.62,31.24)(19.79,31.31)(19.96,31.38)(20.15,31.45)(20.35,31.5)(20.54,31.56)(20.73,31.61)(20.95,31.68)(21.17,31.72)
\path(21.17,31.72)(21.38,31.77)(21.61,31.81)(21.85,31.84)(22.09,31.88)(22.34,31.9)(22.59,31.93)(22.86,31.95)(23.13,31.97)(23.4,31.99)
\path(23.4,31.99)(23.7,31.99)(23.99,32.0)(24.0,32.0)
\path(32.0,32.0)(32.0,32.0)(32.29,31.99)(32.58,31.99)(32.86,31.97)(33.13,31.95)(33.38,31.93)(33.65,31.9)(33.9,31.88)(34.13,31.84)
\path(34.13,31.84)(34.38,31.81)(34.59,31.77)(34.81,31.72)(35.04,31.68)(35.25,31.61)(35.45,31.56)(35.63,31.5)(35.83,31.45)(36.02,31.38)
\path(36.02,31.38)(36.2,31.31)(36.36,31.24)(36.54,31.15)(36.7,31.08)(36.86,31.0)(37.0,30.91)(37.15,30.83)(37.31,30.75)(37.45,30.65)
\path(37.45,30.65)(37.58,30.56)(37.72,30.45)(37.84,30.36)(37.97,30.27)(38.09,30.16)(38.2,30.06)(38.33,29.95)(38.43,29.84)(38.54,29.74)
\path(38.54,29.74)(38.65,29.63)(38.75,29.52)(38.86,29.4)(38.97,29.29)(39.06,29.18)(39.15,29.06)(39.25,28.95)(39.36,28.83)(39.45,28.7)
\path(39.45,28.7)(39.54,28.59)(39.63,28.47)(39.72,28.34)(39.81,28.22)(39.9,28.11)(39.99,28.0)(40.09,27.88)(40.18,27.75)(40.27,27.63)
\path(40.27,27.63)(40.36,27.52)(40.45,27.4)(40.54,27.27)(40.63,27.15)(40.72,27.04)(40.83,26.93)(40.91,26.81)(41.02,26.7)(41.11,26.58)
\path(41.11,26.58)(41.22,26.47)(41.33,26.36)(41.43,26.25)(41.54,26.13)(41.65,26.02)(41.77,25.93)(41.9,25.81)(42.02,25.72)(42.13,25.61)
\path(42.13,25.61)(42.27,25.52)(42.4,25.43)(42.54,25.34)(42.68,25.25)(42.83,25.15)(42.97,25.06)(43.13,24.99)(43.29,24.9)(43.45,24.83)
\path(43.45,24.83)(43.61,24.75)(43.79,24.68)(43.97,24.61)(44.15,24.54)(44.34,24.47)(44.54,24.41)(44.74,24.36)(44.95,24.31)(45.16,24.25)
\path(45.16,24.25)(45.38,24.22)(45.61,24.18)(45.84,24.13)(46.09,24.11)(46.34,24.08)(46.59,24.04)(46.86,24.02)(47.13,24.02)(47.4,24.0)
\path(47.4,24.0)(47.7,24.0)(47.99,24.0)(48.0,24.0)
\drawpath{24.0}{32.0}{32.0}{32.0}
\drawpath{48.0}{24.0}{52.0}{24.0}
\drawpath{4.0}{24.0}{8.0}{24.0}
\drawpath{60.0}{24.0}{64.0}{24.0}
\path(64.0,24.0)(64.0,24.0)(64.29,23.99)(64.58,23.99)(64.86,23.97)(65.12,23.95)(65.38,23.93)(65.65,23.9)(65.9,23.88)(66.13,23.84)
\path(66.13,23.84)(66.37,23.81)(66.59,23.77)(66.81,23.72)(67.04,23.68)(67.25,23.61)(67.44,23.56)(67.63,23.5)(67.83,23.45)(68.01,23.38)
\path(68.01,23.38)(68.19,23.31)(68.37,23.24)(68.54,23.15)(68.69,23.08)(68.86,23.0)(69.01,22.91)(69.16,22.83)(69.3,22.75)(69.44,22.65)
\path(69.44,22.65)(69.58,22.56)(69.72,22.45)(69.84,22.36)(69.97,22.27)(70.08,22.16)(70.2,22.06)(70.33,21.95)(70.44,21.84)(70.55,21.74)
\path(70.55,21.74)(70.66,21.63)(70.76,21.52)(70.87,21.4)(70.97,21.29)(71.06,21.18)(71.16,21.06)(71.26,20.95)(71.36,20.83)(71.44,20.7)
\path(71.44,20.7)(71.54,20.59)(71.62,20.47)(71.72,20.34)(71.8,20.22)(71.9,20.11)(72.0,20.0)(72.08,19.88)(72.18,19.75)(72.26,19.63)
\path(72.26,19.63)(72.36,19.52)(72.44,19.4)(72.54,19.27)(72.62,19.15)(72.73,19.04)(72.83,18.93)(72.91,18.81)(73.01,18.7)(73.12,18.58)
\path(73.12,18.58)(73.23,18.47)(73.33,18.36)(73.44,18.25)(73.55,18.13)(73.66,18.02)(73.77,17.93)(73.9,17.81)(74.01,17.72)(74.13,17.61)
\path(74.13,17.61)(74.26,17.52)(74.41,17.43)(74.54,17.34)(74.68,17.25)(74.83,17.15)(74.98,17.06)(75.12,16.99)(75.29,16.9)(75.44,16.83)
\path(75.44,16.83)(75.62,16.75)(75.79,16.68)(75.97,16.61)(76.16,16.54)(76.34,16.47)(76.54,16.41)(76.73,16.36)(76.94,16.31)(77.16,16.25)
\path(77.16,16.25)(77.38,16.22)(77.61,16.18)(77.84,16.13)(78.08,16.11)(78.33,16.08)(78.59,16.04)(78.86,16.02)(79.12,16.02)(79.41,16.0)
\path(79.41,16.0)(79.69,16.0)(79.98,16.0)(80.0,16.0)
\path(88.0,16.0)(88.0,16.0)(88.29,16.0)(88.58,16.0)(88.86,16.02)(89.12,16.02)(89.38,16.04)(89.65,16.08)(89.9,16.11)(90.13,16.13)
\path(90.13,16.13)(90.37,16.18)(90.59,16.22)(90.81,16.25)(91.04,16.31)(91.25,16.36)(91.44,16.41)(91.63,16.47)(91.83,16.54)(92.01,16.61)
\path(92.01,16.61)(92.19,16.68)(92.37,16.75)(92.54,16.83)(92.69,16.9)(92.86,16.99)(93.01,17.06)(93.16,17.15)(93.3,17.25)(93.44,17.34)
\path(93.44,17.34)(93.58,17.43)(93.72,17.52)(93.84,17.61)(93.97,17.72)(94.08,17.81)(94.2,17.93)(94.33,18.02)(94.44,18.13)(94.55,18.25)
\path(94.55,18.25)(94.66,18.36)(94.76,18.47)(94.87,18.58)(94.97,18.7)(95.06,18.81)(95.16,18.93)(95.26,19.04)(95.36,19.15)(95.44,19.27)
\path(95.44,19.27)(95.54,19.4)(95.62,19.52)(95.72,19.63)(95.8,19.75)(95.9,19.88)(96.0,20.0)(96.08,20.11)(96.18,20.22)(96.26,20.34)
\path(96.26,20.34)(96.36,20.47)(96.44,20.59)(96.54,20.7)(96.62,20.83)(96.73,20.95)(96.83,21.06)(96.91,21.18)(97.01,21.29)(97.12,21.4)
\path(97.12,21.4)(97.23,21.52)(97.33,21.63)(97.44,21.74)(97.55,21.84)(97.66,21.95)(97.77,22.06)(97.9,22.16)(98.01,22.27)(98.13,22.36)
\path(98.13,22.36)(98.26,22.45)(98.41,22.56)(98.54,22.65)(98.68,22.74)(98.83,22.83)(98.98,22.91)(99.12,23.0)(99.29,23.08)(99.44,23.15)
\path(99.44,23.15)(99.62,23.24)(99.79,23.31)(99.97,23.38)(100.16,23.45)(100.34,23.5)(100.54,23.56)(100.73,23.61)(100.94,23.68)(101.16,23.72)
\path(101.16,23.72)(101.38,23.77)(101.61,23.81)(101.84,23.84)(102.08,23.88)(102.33,23.9)(102.59,23.93)(102.86,23.95)(103.12,23.97)(103.41,23.99)
\path(103.41,23.99)(103.69,23.99)(103.98,24.0)(104.0,24.0)
\drawpath{80.0}{16.0}{88.0}{16.0}
\drawpath{104.0}{24.0}{108.0}{24.0}
\thinlines
\drawpath{28.0}{36.0}{28.0}{12.0}
\drawpath{28.0}{12.0}{30.0}{14.0}
\drawpath{84.0}{36.0}{84.0}{12.0}
\drawpath{28.0}{14.0}{30.0}{16.0}
\drawpath{28.0}{16.0}{30.0}{18.0}
\drawpath{28.0}{18.0}{30.0}{20.0}
\drawpath{28.0}{20.0}{30.0}{22.0}
\drawpath{28.0}{22.0}{30.0}{24.0}
\drawpath{28.0}{24.0}{30.0}{26.0}
\drawpath{28.0}{26.0}{30.0}{28.0}
\drawpath{28.0}{28.0}{30.0}{30.0}
\drawpath{28.0}{30.0}{30.0}{32.0}
\drawpath{28.0}{32.0}{30.0}{34.0}
\drawpath{28.0}{34.0}{30.0}{36.0}
\drawpath{84.0}{12.0}{86.0}{14.0}
\drawpath{84.0}{14.0}{86.0}{16.0}
\drawpath{84.0}{16.0}{86.0}{18.0}
\drawpath{84.0}{18.0}{86.0}{20.0}
\drawpath{84.0}{20.0}{86.0}{22.0}
\drawpath{84.0}{22.0}{86.0}{24.0}
\drawpath{84.0}{24.0}{86.0}{26.0}
\drawpath{84.0}{26.0}{86.0}{28.0}
\drawpath{84.0}{28.0}{86.0}{30.0}
\drawpath{84.0}{30.0}{86.0}{32.0}
\drawpath{84.0}{32.0}{86.0}{34.0}
\drawpath{84.0}{34.0}{86.0}{36.0}
\drawvector{18.0}{28.0}{6.0}{1}{0}
\drawvector{38.0}{28.0}{6.0}{-1}{0}
\drawvector{70.0}{20.0}{6.0}{-1}{0}
\drawvector{98.0}{20.0}{6.0}{1}{0}
\drawcenteredtext{28.0}{6.0}{before}
\drawcenteredtext{84.0}{6.0}{after}
\drawpath{4.0}{24.0}{52.0}{24.0}
\drawpath{60.0}{24.0}{108.0}{24.0}
\end{picture}
\setlength{\unitlength}{1mm}
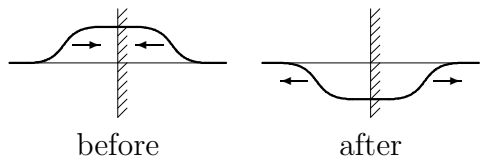
\caption{Soliton with mirror antisoliton satisfies Neumann
condition \eqref{c1}.\label{fig1}}
}

The first affine Toda theory to be studied in the presence of 
integrable boundaries was the sine-Gordon theory 
\cite{Gho,Mac94,Sal94}.
The first necessary step to extend this to higher affine Toda theories
is to find those boundary conditions which preserve
integrability. This was done by the Durham group
in \cite{Cor94a,Cor95} by studying the first
higher-spin integrals of motion and then more systematically
in \cite{Bow95} by constructing the generating functional for all
higher-spin integrals of motion using a Lax pair construction.
They find integrable boundary conditions of
the form
\begin{equation}
\partial _x\bphi|_{\text{boundary}}=\left.\sum_{i=0}^n\ A_i\,
\sqrt{{2\eta _i}/{|\aroot_i|^2}}\,\aroot_i\ e^{\aroot_i\cdot 
\bphi/2}\right|_{\text{boundary}}.  \label{cbc}
\end{equation}
The coefficients $A_i$ specify the boundary condition.
In the sine-Gordon model the $A_i$ are allowed to take
arbitrary values but in all other affine Toda theories 
either all the $A_i$ must be zero (the Neumann boundary condition)
or else most of the
$A_i$ are restricted to be $\pm 1$ if integrability is to
be preserved \cite{Bow95}. 

The existence of an infinite set of integrals of motion in involution
is one hallmark of integrability, the existence of soliton solutions
is another. In \cite{Del98} we showed that it is indeed possible 
to construct soliton solutions
satisfying boundary conditions of the form \eqref{cbc}. We used a
method of mirror images. Let us recall our method in the simplest 
case of the Neumann boundary condition $\partial_x\bphi=0$ at $x=0$. 
Under parity reversal $\bphi(x)\rightarrow\tilde{\bphi}(x)=\bphi(-x)$
a soliton with center of mass $X$ moving with velocity $v$ is
transformed into an antisoliton with center of mass $-X$ and
velocity $-v$. Therefore if $\bphi$ describes a multisoliton
configuration in which every soliton is paired with an oppositely
moving antisoliton centered around the parity-transformed point, then
$\bphi$ is invariant under parity, 
\begin{equation}\label{c1}
\bphi(-x)=\bphi(x)~~\text{ and thus }~~\partial_x\bphi(x=0)=0.
\end{equation}
Viewed on the left half line such a configuration describes solitons
hitting the boundary and coming back as antisolitons. This is described
schematically in figure \ref{fig1}.

\FIGURE{
\setlength{\unitlength}{0.6mm}
\begin{picture}(112,34)
\thicklines
\path(8.0,12.0)(8.0,12.0)(8.29,12.0)(8.57,12.0)(8.86,12.02)(9.12,12.02)(9.38,12.04)(9.64,12.08)(9.89,12.11)(10.13,12.13)
\path(10.13,12.13)(10.37,12.18)(10.6,12.22)(10.81,12.25)(11.04,12.31)(11.25,12.36)(11.45,12.41)(11.63,12.47)(11.82,12.54)(12.02,12.61)
\path(12.02,12.61)(12.2,12.68)(12.37,12.75)(12.54,12.83)(12.7,12.9)(12.86,12.99)(13.01,13.06)(13.15,13.15)(13.3,13.25)(13.45,13.34)
\path(13.45,13.34)(13.57,13.43)(13.71,13.52)(13.85,13.61)(13.96,13.72)(14.09,13.81)(14.2,13.93)(14.32,14.02)(14.44,14.13)(14.54,14.25)
\path(14.54,14.25)(14.65,14.36)(14.76,14.47)(14.87,14.58)(14.96,14.7)(15.06,14.81)(15.15,14.93)(15.26,15.04)(15.36,15.15)(15.45,15.27)
\path(15.45,15.27)(15.54,15.4)(15.62,15.52)(15.71,15.63)(15.8,15.75)(15.89,15.88)(15.98,16.0)(16.09,16.11)(16.18,16.22)(16.27,16.34)
\path(16.27,16.34)(16.36,16.47)(16.45,16.59)(16.54,16.7)(16.63,16.83)(16.72,16.95)(16.83,17.06)(16.91,17.18)(17.02,17.29)(17.11,17.4)
\path(17.11,17.4)(17.22,17.52)(17.33,17.63)(17.43,17.74)(17.54,17.84)(17.65,17.95)(17.77,18.06)(17.9,18.16)(18.02,18.27)(18.13,18.36)
\path(18.13,18.36)(18.27,18.45)(18.4,18.56)(18.54,18.65)(18.68,18.74)(18.83,18.83)(18.97,18.91)(19.13,19.0)(19.29,19.08)(19.45,19.15)
\path(19.45,19.15)(19.61,19.24)(19.79,19.31)(19.97,19.38)(20.15,19.45)(20.34,19.5)(20.54,19.56)(20.74,19.61)(20.95,19.68)(21.16,19.72)
\path(21.16,19.72)(21.38,19.77)(21.61,19.81)(21.84,19.84)(22.09,19.88)(22.34,19.9)(22.59,19.93)(22.86,19.95)(23.13,19.97)(23.4,19.99)
\path(23.4,19.99)(23.7,19.99)(23.99,20.0)(24.0,20.0)
\path(32.0,20.0)(32.0,20.0)(32.29,20.0)(32.58,20.0)(32.86,20.02)(33.13,20.02)(33.38,20.04)(33.65,20.08)(33.9,20.11)(34.13,20.13)
\path(34.13,20.13)(34.38,20.18)(34.59,20.22)(34.81,20.25)(35.04,20.31)(35.25,20.36)(35.45,20.41)(35.63,20.47)(35.83,20.54)(36.02,20.61)
\path(36.02,20.61)(36.2,20.68)(36.36,20.75)(36.54,20.83)(36.7,20.9)(36.86,20.99)(37.0,21.06)(37.15,21.15)(37.31,21.25)(37.45,21.34)
\path(37.45,21.34)(37.58,21.43)(37.72,21.52)(37.84,21.61)(37.97,21.72)(38.09,21.81)(38.2,21.93)(38.33,22.02)(38.43,22.13)(38.54,22.25)
\path(38.54,22.25)(38.65,22.36)(38.75,22.47)(38.86,22.58)(38.97,22.7)(39.06,22.81)(39.15,22.93)(39.25,23.04)(39.36,23.15)(39.45,23.27)
\path(39.45,23.27)(39.54,23.4)(39.63,23.52)(39.72,23.63)(39.81,23.75)(39.9,23.88)(39.99,24.0)(40.09,24.11)(40.18,24.22)(40.27,24.34)
\path(40.27,24.34)(40.36,24.47)(40.45,24.59)(40.54,24.7)(40.63,24.83)(40.72,24.95)(40.83,25.06)(40.91,25.18)(41.02,25.29)(41.11,25.4)
\path(41.11,25.4)(41.22,25.52)(41.33,25.63)(41.43,25.74)(41.54,25.84)(41.65,25.95)(41.77,26.06)(41.9,26.16)(42.02,26.27)(42.13,26.36)
\path(42.13,26.36)(42.27,26.45)(42.4,26.56)(42.54,26.65)(42.68,26.74)(42.83,26.83)(42.97,26.91)(43.13,27.0)(43.29,27.08)(43.45,27.15)
\path(43.45,27.15)(43.61,27.24)(43.79,27.31)(43.97,27.38)(44.15,27.45)(44.34,27.5)(44.54,27.56)(44.74,27.61)(44.95,27.68)(45.16,27.72)
\path(45.16,27.72)(45.38,27.77)(45.61,27.81)(45.84,27.84)(46.09,27.88)(46.34,27.9)(46.59,27.93)(46.86,27.95)(47.13,27.97)(47.4,27.99)
\path(47.4,27.99)(47.7,27.99)(47.99,28.0)(48.0,28.0)
\drawpath{4.0}{12.0}{8.0}{12.0}
\drawpath{24.0}{20.0}{32.0}{20.0}
\drawpath{48.0}{28.0}{52.0}{28.0}
\thinlines
\drawpath{28.0}{30.0}{28.0}{10.0}
\drawpath{4.0}{12.0}{52.0}{12.0}
\drawpath{28.0}{10.0}{30.0}{12.0}
\drawpath{28.0}{12.0}{30.0}{14.0}
\drawpath{28.0}{14.0}{30.0}{16.0}
\drawpath{28.0}{16.0}{30.0}{18.0}
\drawpath{28.0}{18.0}{30.0}{20.0}
\drawpath{28.0}{20.0}{30.0}{22.0}
\drawpath{28.0}{22.0}{30.0}{24.0}
\drawpath{28.0}{24.0}{30.0}{26.0}
\drawpath{28.0}{26.0}{30.0}{28.0}
\drawpath{28.0}{28.0}{30.0}{30.0}
\thicklines
\path(64.0,12.0)(64.0,12.0)(64.29,12.0)(64.58,12.0)(64.86,12.02)(65.12,12.02)(65.38,12.04)(65.65,12.08)(65.9,12.11)(66.13,12.13)
\path(66.13,12.13)(66.37,12.18)(66.59,12.22)(66.81,12.25)(67.04,12.31)(67.25,12.36)(67.44,12.41)(67.63,12.47)(67.83,12.54)(68.02,12.61)
\path(68.02,12.61)(68.19,12.68)(68.37,12.75)(68.54,12.83)(68.69,12.9)(68.86,12.99)(69.01,13.06)(69.16,13.15)(69.3,13.25)(69.44,13.34)
\path(69.44,13.34)(69.58,13.43)(69.72,13.52)(69.84,13.61)(69.97,13.72)(70.08,13.81)(70.2,13.93)(70.33,14.02)(70.44,14.13)(70.55,14.25)
\path(70.55,14.25)(70.66,14.36)(70.76,14.47)(70.87,14.58)(70.97,14.7)(71.06,14.81)(71.16,14.93)(71.26,15.04)(71.36,15.15)(71.44,15.27)
\path(71.44,15.27)(71.54,15.4)(71.62,15.52)(71.72,15.63)(71.8,15.75)(71.9,15.88)(71.98,16.0)(72.08,16.11)(72.18,16.22)(72.27,16.34)
\path(72.27,16.34)(72.36,16.47)(72.44,16.59)(72.54,16.7)(72.63,16.83)(72.72,16.95)(72.83,17.06)(72.91,17.18)(73.02,17.29)(73.11,17.4)
\path(73.11,17.4)(73.22,17.52)(73.33,17.63)(73.44,17.74)(73.55,17.84)(73.66,17.95)(73.77,18.06)(73.9,18.16)(74.02,18.27)(74.13,18.36)
\path(74.13,18.36)(74.27,18.45)(74.41,18.56)(74.54,18.65)(74.68,18.74)(74.83,18.83)(74.97,18.91)(75.13,19.0)(75.29,19.08)(75.44,19.15)
\path(75.44,19.15)(75.61,19.24)(75.79,19.31)(75.97,19.38)(76.16,19.45)(76.34,19.5)(76.54,19.56)(76.74,19.61)(76.94,19.68)(77.16,19.72)
\path(77.16,19.72)(77.38,19.77)(77.61,19.81)(77.84,19.84)(78.08,19.88)(78.33,19.9)(78.59,19.93)(78.86,19.95)(79.13,19.97)(79.41,19.99)
\path(79.41,19.99)(79.69,19.99)(79.99,20.0)(80.0,20.0)
\path(88.0,20.0)(88.0,20.0)(88.29,20.0)(88.58,20.0)(88.86,20.02)(89.13,20.02)(89.38,20.04)(89.65,20.08)(89.9,20.11)(90.13,20.13)
\path(90.13,20.13)(90.38,20.18)(90.59,20.22)(90.81,20.25)(91.04,20.31)(91.25,20.36)(91.44,20.41)(91.63,20.47)(91.83,20.54)(92.02,20.61)
\path(92.02,20.61)(92.19,20.68)(92.36,20.75)(92.54,20.83)(92.69,20.9)(92.86,20.99)(93.0,21.06)(93.16,21.15)(93.3,21.25)(93.44,21.34)
\path(93.44,21.34)(93.58,21.43)(93.72,21.52)(93.84,21.61)(93.97,21.72)(94.08,21.81)(94.2,21.93)(94.33,22.02)(94.44,22.13)(94.55,22.25)
\path(94.55,22.25)(94.66,22.36)(94.75,22.47)(94.86,22.58)(94.97,22.7)(95.06,22.81)(95.16,22.93)(95.25,23.04)(95.36,23.15)(95.44,23.27)
\path(95.44,23.27)(95.54,23.4)(95.63,23.52)(95.72,23.63)(95.8,23.75)(95.9,23.88)(95.99,24.0)(96.08,24.11)(96.18,24.22)(96.27,24.34)
\path(96.27,24.34)(96.36,24.47)(96.44,24.59)(96.54,24.7)(96.63,24.83)(96.72,24.95)(96.83,25.06)(96.91,25.18)(97.02,25.29)(97.11,25.4)
\path(97.11,25.4)(97.22,25.52)(97.33,25.63)(97.44,25.74)(97.55,25.84)(97.66,25.95)(97.77,26.06)(97.9,26.16)(98.02,26.27)(98.13,26.36)
\path(98.13,26.36)(98.27,26.45)(98.41,26.56)(98.54,26.65)(98.68,26.74)(98.83,26.83)(98.97,26.91)(99.13,27.0)(99.29,27.08)(99.44,27.15)
\path(99.44,27.15)(99.61,27.24)(99.79,27.31)(99.97,27.38)(100.16,27.45)(100.34,27.5)(100.54,27.56)(100.74,27.61)(100.94,27.68)(101.16,27.72)
\path(101.16,27.72)(101.38,27.77)(101.61,27.81)(101.84,27.84)(102.08,27.88)(102.33,27.9)(102.59,27.93)(102.86,27.95)(103.13,27.97)(103.41,27.99)
\path(103.41,27.99)(103.69,27.99)(103.99,28.0)(104.0,28.0)
\drawpath{60.0}{12.0}{64.0}{12.0}
\drawpath{80.0}{20.0}{88.0}{20.0}
\drawpath{104.0}{28.0}{108.0}{28.0}
\thinlines
\drawpath{84.0}{30.0}{84.0}{10.0}
\drawpath{60.0}{12.0}{108.0}{12.0}
\drawpath{84.0}{10.0}{86.0}{12.0}
\drawpath{84.0}{12.0}{86.0}{14.0}
\drawpath{84.0}{14.0}{86.0}{16.0}
\drawpath{84.0}{16.0}{86.0}{18.0}
\drawpath{84.0}{18.0}{86.0}{20.0}
\drawpath{84.0}{20.0}{86.0}{22.0}
\drawpath{84.0}{22.0}{86.0}{24.0}
\drawpath{84.0}{24.0}{86.0}{26.0}
\drawpath{84.0}{26.0}{86.0}{28.0}
\drawpath{84.0}{28.0}{86.0}{30.0}
\drawcenteredtext{28.0}{6.0}{before}
\drawcenteredtext{84.0}{6.0}{after}
\drawvector{18.0}{16.0}{6.0}{1}{0}
\drawvector{38.0}{24.0}{6.0}{-1}{0}
\drawvector{70.0}{16.0}{6.0}{-1}{0}
\drawvector{98.0}{24.0}{6.0}{1}{0}
\end{picture}
\setlength{\unitlength}{1mm}
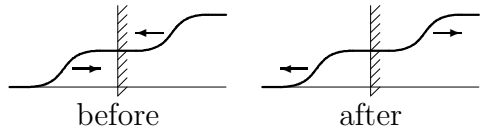
\caption{Soliton with mirror soliton satisfies the new
condition \eqref{c2}.\label{fig2}}
}
In this letter we want to find a boundary condition which ensures 
that solitons come back as solitons after reflection. Thus we want
a condition which selects those multisoliton solutions on the whole line
in which each soliton is paired with a mirror soliton rather than
a mirror antisoliton, see figure \ref{fig2}. 
Using the observation that complex conjugation
transforms a soliton configuration into an antisoliton configuration
we see that the appropriate modification of the condition \eqref{c1}
is
\begin{equation}\label{c2}
\bphi^*(-x)=\bphi(x)+2\pi i \weight~~\text{with}~~
\weight\in\text{coweight lattice of }\lie.
\end{equation}
The freedom is due to the fact that adding $2\pi i$ times a coweight
to a soliton solution gives an equivalent soliton solution.
The boundary condition corresponding to \eqref{c2} is
\begin{equation}\label{nbc}
\left.\frac{\imaginary{\bphi}}{2\pi}\right|_{\text{boundary}}
\in\text{coweight lattice of }\lie~~~
\text{and}~~~\left.\partial_x\real{\bphi}\right|_{\text{boundary}}=0,
\end{equation}
It is a combination of a Dirichlet boundary condition for the
imaginary part of the field and a Neumann boundary condition for the
real part.

In the sine-Gordon model the field is purely imaginary
(in our conventions). Thus in that case the boundary condition 
\eqref{nbc} is simply a Dirichlet condition which was known to 
be integrable. In general affine Toda theories one has to include
complex field configurations in order to have solitons. The
boundary condition \eqref{nbc}
is the integrable generalization of the Dirichlet condition.
Note that unlike in the sine-Gordon case there is no free parameter in the 
boundary condition \eqref{nbc}.

It is intuitively clear that the new boundary condition \eqref{nbc}
preserves exactly half of the infinitely many conserved charges of
the theory on the whole line: when a right moving
soliton is about to leave the
left half line and to carry some charge away with it,
a left moving soliton is moving onto the left half line. If a left
moving soliton carries the same amount of charge as a right moving
one, then the amount of charge on the left half line will be conserved.
This is so exactly for half of all the charges.
In the following sections we will go beyond this intuitive argument
and construct a generating functional for the integrals of motion.
We will use a generalization of Sklyanin's formalism \cite{Skl87},
which we find conceptually slightly simpler than the approach of 
\cite{Bow95}, and which works for intervals as well as the half line. We
will also review the Durham boundary conditions in this formalism.
The reason why the boundary condition \eqref{nbc} was not discovered in
\cite{Bow95} is that there the field was always
restricted to be purely real.

\section{Sklyanin's formalism for integrable field theories with boundaries
\label{review}}

Sklyanin's original presentation of his method in his two page paper
\cite{Skl87} is very succinct and clear. We present it here in a slightly
generalized form needed for affine Toda theory.
The method applies to spin chains as well as to
field theories in two dimensions, we concentrate on the later.

It is assumed that 
there exists a pair of functions $a_x(\lam)$ and $a_t(\lam)$ which
take their values in some Lie algebra $\lie$ and which depend
on the fields of the theory, their conjugate momenta, 
and on an extra real parameter $\lam$, so that the
equations of motion of the field theory on 
the whole line are equivalent to the Lax pair equation
\begin{equation}\label{lax}
\left[\partial_x + a_x(\lam), \partial_t + a_t(\lam)\right]=0~~~
\forall\lam.
\end{equation}
The utility of the extra parameter $\lam$, called spectral parameter, 
will become apparent later. 
$a_x$ and $a_t$ depend on $x$ and $t$ only implicitly through their
dependence on the fields. $\partial_x$ and $\partial_t$ denote total
differentiation with respect to the space or time variable.
A concrete example of a Lax pair
will be given for affine Toda
theory in \eq{pair1}. If $a_x$ and $a_t$ are thought of as the components of a 
connection then \eq{lax} is the zero curvature condition for this
connection.

It is furthermore assumed that the canonical Poisson brackets for the
fields give rise to Poisson brackets for the functions $a_x(\lam;x)$ 
of the form
\begin{equation}\label{apoisson}
\left\{\one{a}_x(\lam_1;x),\two{a}_x(\lam_2;y)\right\}=
\delta(x-y)\left[\ot{r}(\lam_1/\lam_2),
\one{a}_x(\lam_1;x)+\two{a}_x(\lam_2;y)\right].
\end{equation}
This is to be understood as an equation in $(\lie\oplus 1)\otimes
(\lie\oplus 1)$.
We used the shorthand notation $\one{a}_x=a_x\otimes 1$ and
$\two{a}_x=1\otimes a_x$. $\ot{r}$ is an element of 
$\lie\otimes \lie$
which satisfies the classical Yang-Baxter equation. It is assumed to
be constant and independent of the fields. 
In \cite{Skl87} Sklyanin
requires that the $r$-matrix satisfies $r(\lam)=-r(1/\lam)$. We drop
this requirement because it is not essential and 
is not satisfied by the $r$-matrix \eqref{r1} of affine Toda theory.

We now restrict the theory to an interval $[x_-, x_+]$.
We define the "transfer matrix" 
$T(\lam)$
as the path ordered exponentials of $a_x$ from $x_-$ to $x_+$,
\begin{equation}
T(\lam)=P\exp\left(\int_{x_-}^{x_+}a_x(\lam,x)\,dx\right).
\end{equation}
The path ordered exponential is defined so that the operators at points
nearer to $x_+$ are further to the left. These $T(\lam)$
satisfy
\begin{align}
\partial_{x_+}T(\lam)&=a_x(\lam;x_+)T(\lam),\\
\partial_{t}T(\lam)&=a_t(\lam;x_+)T(\lam)-
T(\lam)a_t(\lam;x_-).
\end{align}
The Poisson bracket relation \eqref{apoisson} leads to 
\begin{equation}\label{tpoisson}
\left\{\one{T}(\lam_1),\two{T}(\lam_2)\right\}=
\left[\ot{r}(\lam_1/\lam_2),\one{T}(\lam_1)\two{T}(\lam_2)\right].
\end{equation}
In practice one will often work in a specific chosen representation
of $\lie$. We will use the same symbols for Lie algebra valued
objects and for the corresponding representation matrices.

Following Sklyanin one can prove the following proposition by
straightforward calculation.

\begin{proposition}\label{prop}
Let the Lax connection $a_\mu(\lam)$, the transfer matrix $T(\lam)$
and the $r$-matrix $r(\lam)$ be defined as above, but now written
as matrices in some chosen representation.
Let $\invo$ be a linear or anti-linear involutive anti-automorphism of the
Lie algebra $\lie$.
Let $K_\pm(\lam)$ be two matrix valued functions of the spectral parameter
$\lam$, not depending on the fields or space-time, satisfying the classical
reflection equations
\begin{align}\label{refl}
0=&\one{K}_+(\lam_1)\two{K}_+(\lam_2)r(\lam_1/\lam_2)+
r^{\invo_1\invo_2}(\lam_2/\lam_1)\one{K}_+(\lam_1)\two{K}_+(\lam_2)
\nonumber\\
&+\one{K}_+(\lam_1) r^{\invo_2}(\lam_1 \lam_2)\two{K}_+(\lam_2)+
\two{K}_+(\lam_2) r^{\invo_1}(1/(\lam_1\lam_2))\one{K}_+(\lam_1),
\\\label{mrefl}
0=&\one{K}_-(\lam_1)\two{K}_-(\lam_2)r^{\invo_1\invo_2}(\lam_1/\lam_2)+
r(\lam_2/\lam_1)\one{K}_-(\lam_1)\two{K}_-(\lam_2)
\nonumber\\
&+\one{K}_-(\lam_1) r^{\invo_1}(\lam_1 \lam_2)\two{K}_-(\lam_2)+
\two{K}_-(\lam_2) r^{\invo_2}(1/(\lam_1\lam_2))\one{K}_-(\lam_1),
\end{align}
where by $r^{\invo_1}$ we mean the result of acting with the involution
$\invo$ on the first factor of $r$ and
similarly for $r^{\invo_2}$ and $r^{\invo_1\invo_2}$.
Then: 

a) The function
\begin{equation}\label{trans}
\ct(u)=T(\lam)K_-(1/\lam)T^{\invo}(1/\lam)
\end{equation}
obeys the Poisson bracket relation
\begin{multline}\label{poisson}
\left\{\one{\ct}(\lam_1),\two{\ct}(\lam_2)\right\}=
r(\lam_1/\lam_2)\one{\ct}(\lam_1)\two{\ct}(\lam_2)+
\one{\ct}(\lam_1)\two{\ct}(\lam_2)r^{\invo_1\invo_2}(\lam_2/\lam_1)\\
+\two{\ct}(\lam_2) r^{\invo_2}(\lam_1 \lam_2)\one{\ct}(\lam_1)+
\one{\ct}(\lam_1) r^{\invo_1}(1/(\lam_1\lam_2))\two{\ct}(\lam_2).
\end{multline}

b) The quantities
\begin{equation}\label{cct}
\cct(\lam)=\tr\left(K_+(\lam)\ct(\lam)\right)=
\tr\left(K_+(\lam)T(\lam)K_-(1/\lam)T^{\invo}(1/\lam)\right)
\end{equation}
are in involution for any two values of the spectral parameter,
\begin{equation}
\left\{\cct(\lam_1),\cct(\lam_2)\right\}=0,~~~\forall 
\lam_1,\lam_2\in\reals.
\end{equation}

c) If the the boundary conditions on the fields at $x_-$ and $x_+$ are such
that they lead to
\begin{gather}\label{ibc}
K_+(\lam)a_t(\lam;x_+)+a_t^\invo(1/\lam;x_+)K_+(\lam)=0,\\\label{mbc}
K_-(\lam)a_t^\invo(\lam;x_-)+a_t(1/\lam;x_-)K_-(\lam)=0,
\end{gather}
then the quantities $\cct(\lam)$ defined in \eq{cct} are time conserved,
\begin{equation}
\frac{d}{dt}\,\cct(\lam)=0,~~~\forall\lam\in\reals.
\end{equation}
\end{proposition}

One obtains Sklyanin's result \cite{Skl87} from this by choosing
the anti-automorphism $\invo$ to be the antipode
\footnote{The antipode acts on Lie
algebra elements by multiplying them by $-1$ and thus acts
on group elements by sending them to their inverse.}
and by assuming
that $r(1/\lam)=-r(\lam)$. 
Sklyanin \cite{Skl87} writes his formulas 
in terms of an additive spectral parameter $u$ rather than the 
multiplicative spectral parameter $\lam=e^u$ which we use.
The generalization of allowing antiautomorphisms other than the
antipode was suggested to us by a similar generalization of the
quantum reflection equation \cite{Kul}. Something similar has also been found
in \cite{Chen}.

The proposition implies that
we can take $\cct$ as the generating functional for the integrals of
motion in involution. If $\cct(\lam)$ has a non-trivial 
dependence on $\lam$ so that it generates an infinite number of 
independent integrals of motion, the field theory on the interval 
$[x_-,x_+]$ with equations of motion
given by \eqref{lax} and boundary conditions given by \eqref{ibc}
is integrable.

\section{Lax pair and $r$-matrix for affine Toda theory}

The Lax pair for the Toda theory based on an untwisted
\footnote{A restriction which we make only for compactness of
presentation} affine Lie algebra
$\lie^{(1)}$ takes values in the finite dimensional Lie algebra $\lie$,
\cite{Oli85}
\begin{align}\label{pair1}
a_t(\lam)&=\half\partial_x\bphi\cdot \bH
+\sum_{i=0}^n\sqrt{\kac_i\al_i^2/8}
\left(\lam E_{\al_i}-1/\lam E_{-\al_i}\right)e^{\al_i\cdot\bphi/2},
\nonumber\\
a_x(\lam)&=\half\partial_t\bphi\cdot \bH
+\sum_{i=0}^n\sqrt{\kac_i\al_i^2/8}
\left(\lam E_{\al_i}+1/\lam E_{-\al_i}\right)e^{\al_i\cdot\bphi/2}.
\end{align}
Substituting the expressions \eqref{pair1} into the zero curvature
condition \eqref{lax} and using the Lie bracket relations
\begin{align}
[\bH,E_\al]&=\aroot\,E_\al,&[E_\al,E_\al]&=0,&
[E_\al,E_{-\al}]&=\frac{2}{\al^2}\,\aroot\cdot\bH.
\end{align}
reproduces the equations of motion \eqref{phieom}.

The canonical equal-time Poisson bracket relations
\begin{equation}
\{\phi_a(x,t),\partial_t\phi_b(y,t)\}=\delta_{ab}\delta(x-y)
\end{equation}
lead to Poisson brackets for the $a_x$ of the form
\eqref{apoisson} with the $r$-matrix \cite{Oli85}
\begin{align}\label{r1}
r(\lam)\propto\frac{\lam^h+1}{\lam^h-1}\sum_{a=1}^n \,H_a\otimes H_a
&-\frac{2}{1-\lam^h}\sum_{\al>0}\frac{\al^2}{2}\lam^{l(\al)}
E_\al\otimes E_{-\al}\nonumber\\
&+\frac{2}{1-\lam^{-h}}\sum_{\al>0}\frac{\al^2}{2}\lam^{-l(\al)}
E_{-\al}\otimes E_{\al},
\end{align}
where the generators are chosen so that
$\tr(H_a\,H_b)=\delta_{ab}$ and 
$\tr(E_\al E_{-\beta})=2 \delta_{\al\beta}/\al^2$.

\section{Durham boundary conditions}

We can obtain the Durham boundary conditions \eqref{cbc}
from the Sklyanin formalism by choosing the anti-automorphism
$\invo$ as transposition,
acting on the Lie algebra generators as
$E_\al^t=E_{-\al}$ and $H^t=H$.
The condition \eqref{ibc} for $K_+$ becomes
\begin{equation}\label{keq}
\left[K_+(\lam),\half\partial_x\bphi\cdot \bH\right]_+
+\left[K_+(\lam),\sum_{i=0}^n\sqrt{\kac_i\al_i^2/8}
\left(\lam E_{\al_i}-1/\lam E_{-\al_i}\right)e^{\al_i\cdot\bphi/2}
\right]_-=0
\end{equation}
at $x=x_+$.
This is exactly the same as equation (3.14) in \cite{Bow95} if one
makes the identification $K_+={\cal{K}}^{-1}$ and uses
that $\partial B/\partial\bphi=-\partial_x\bphi$. 
In \cite{Bow95} this equation was derived
by a different formalism in which the Lax pair condition gives
both the equations of motion and the boundary conditions.
The results of the Durham group show that there exists
a $K_+(\lam)$ satisfying \eq{keq} only if the boundary
condition is of the form \eq{cbc} with very restricted choice
for the coefficients $A_i$ and that such a
$K_+(\lam)$ automatically satisfies the reflection equation
\eqref{refl} (Equation (4.8) in \cite{Bow95} is equivalent to our
\eq{refl} because the Toda $r$-matrix has the property that
$r^{t_1t_2}(\lam)=-r(1/\lam)$).
The simplest solution is of course $K_+(\lam)=1$ which solves
\eqref{keq} when the boundary condition is $\partial_x\bphi(x_+)=0$.

If $K_+(\lam)$ is a solution of \eq{ibc} then $K_-(\lam)=K_+(-1/\lam)$
is a solution of \eq{mbc} with the same boundary condition. 
This is so because $a_t^t(-\lam)=a_t(\lam)$.
This $K_-(\lam)$ satisfies the reflection equation \eqref{mrefl}
because $r^{t_1t_2}(\lam)=-r(1/\lam)$.
Thus affine Toda theory is integrable on any interval $[x_-,x_+]$ with
boundary conditions of the form \eqref{cbc} at both ends, where
the coefficients $A_i$ do not have to be the same at the two ends
(but of course independently have to obey the restrictions of 
\cite{Bow95}).
The Durham group dealt only with the
half line which can be obtained by choosing $K_-=1$ and taking the limit 
$x_-\rightarrow -\infty$.

\section{The soliton-preserving boundary conditions}

We will now choose the anti-automorphism $\invo$ to be hermitian
conjugation $\dagger$, defined by its action on the
Lie algebra generators
$E_\al^\dagger=E_{-\al}$, and $H_a^\dagger=H_a$.
It differs from transposition only in that
any coefficients of Lie algebra generators will be complex conjugated
and thus in particular $(\bphi\cdot\bH)^\dagger=\bphi^*\cdot\bH$.
The condition \eqref{ibc} of compatibility of the $K$-matrix
with the Lax connection therefore reads
\begin{align}\label{ke2}
0&=\left[K_+(\lam),\half\partial_x\real{\bphi}\cdot H\right]_+
+\left[K_+(\lam),\sum_{i=0}^n\sqrt{\frac{\kac_i\al_i^2}{8}}
\left(\lam E_{\al_i}-\frac{1}{\lam} E_{-\al_i}\right)
\real{e^{\al_i\cdot\bphi/2}}
\right]_-\\\label{ke3}
0&=\left[K_+(\lam),\half\partial_x\imaginary{\bphi}\cdot H\right]_-
+\left[K_+(\lam),\sum_{i=0}^n\sqrt{\frac{\kac_i\al_i^2}{8}}
\left(\lam E_{\al_i}-\frac{1}{\lam} E_{-\al_i}\right)
\imaginary{e^{\al_i\cdot\bphi/2}}
\right]_+
\end{align}
at $x=x_+$.
If one imposes our new boundary condition \eqref{nbc} then these
equations are solved by $K_+=1$ which is also a solution of the
reflection equation \eqref{refl}. This proves integrability of
the condition \eqref{nbc}.

Also the $K$-matrices found by the Durham group can solve these equations
but only if the boundary conditions are stronger than the conditions
\eqref{cbc} which were already known to be integrable.

\section{Discussion\label{discussion}}

We have seen that the Sklyanin formalism can be generalized by replacing
the antipode by an arbitrary linear or antilinear antiautomorphism.
By choosing this antiautomorphism to be transposition and using the Durham
K-matrices \cite{Bow95} one obtains the known boundary conditions
\eqref{cbc}. By choosing hermitian conjugation instead and choosing
$K=1$ one obtains the new boundary condition \eqref{nbc}.

We did not explicitly extract \textit{local} integrals of motion
in involution from the generating functional $\cct(\lam)$. This might 
perhaps be done by
extending the abelianization formalism of Olive and Turok \cite{Oli85}
to the situation with boundaries. Intuitively we expect the charges
carried by the solitons to take the same values as on the whole line,
determined by Freeman \cite{Fre}.

We are not claiming to have found all integrable boundary conditions.
For example it is probably possible to find more integrable boundary conditions
involving the time derivative of the field by using gauge transformed
Lax pairs as in \cite{Bow95b}.

Gandenberger \cite{Gan98} has successfully determined the soliton-conjugating
quantum reflection matrices for the Durham boundary conditions \eqref{cbc}
for $\lie=sl(3)$. We expect that it will be similarly possible to
find the soliton-preserving
quantum reflection matrices for our boundary condition \eqref{nbc}
from the solutions of the reflection equation given in
\cite{deV}.

\acknowledgments{I would like to thank Ed Corrigan,
Rafael Nepomechie and Petr Kulish for discussions during the 
Euroconference on "New Symmetries in Statistical Mechanics and 
Condensed Matter Physics" funded by TMR contract number
ERBFMMACT970283. This work was supported in part
by the European Commission TMR Network, 
contract number FMRX-CT96-0012.
I thank the EPSRC for an advanced research fellowship.
}

\appendix

\end{document}